\documentclass{elcopy}
\usepackage{natbib}
\usepackage{amssymb}
\usepackage{graphics}
\begin{document}
\begin{frontmatter}

\title{Towards the Typology of Elections at Russia}
\author[ibf,defakto]{Michael G.Sadovsky\corauthref{cor1}}
\corauth[cor1]{To whom the correspondence should be addressed}
\address{660036 Russia, Krasnoyarsk, P.O.Box 8737}
\address[ibf]{Institute of computational modelling of RAS}
\ead{msad@icm.krasn.ru}
\address[defakto]{``Gorod" Producer Company}

\author[saa]{Alexander A.Gliskov}
\address[saa]{Krasnoyarsk regional Chamber of Attorney, barrister}
\ead{gliskov@yandex.ru}

\begin{abstract}
A distinction in reasons and motives for choosing a particular
political leader establishes the key difference between older and
young democracy. The former is based on electoral history, while
the latter is based on feelings and personal attitude. Besides, a
comparatively abundant number of political figures (persons or
parties and associations) is specific for young democracies. The
problem of a reference votes' distribution is analyzed. Lefevbre's
theory of a reflexive control is supposed to make the basis for
indifferent choice of political figures. This theory yields a
golden section split of votes (or the series of Fibonacci numbers,
for the case of multiple choice). A typology of political
campaigns based on this theory is proposed. A proximity of ratings
of competing persons means the highest electoral tension, a
leadership of a person means a high level of mobilization; a
neutral situation corresponds to Fibonacci numbers distribution of
votes.
\end{abstract}

\end{frontmatter}

\section{Introduction}\label{1}
Democratic nations exhibit rather stable, apparent and efficient
pattern of public life. This pattern results from a rather long
electoral history of the nations. Young democracies, which
appeared due to a communist system collapse have to run through
the same experience in considerably shorter period of time. Free,
direct and independent elections constitute a key issue of any
democratic regime. As a rule, an effective democracy exhibits a
considerably short list of political parties, which take part in
the social and political life. On the contrary, young democracies
usually exhibit an extended list of such entities, in comparison
to the elder ones.

Not discussing the origin and current state of political
situations in younger democracies in this paper, lets us instead
concentrate on the question concerning the interpretation of the
elections results in the situation when an abundance of political
parties taking part in a race exists. A main objective of any
elective system is to establish and legitimatize an authority
responsible for a nation, or for a region within a country. That
latter is determined by the elections results. To get a clear and
comprehensive ideas about the legitimate power, one must
understand pretty well what kind of elections results should be
considered as typical, proper and indicative from the public point
of view.

Here we propose an approach to determine such ``reference"
distribution in the election rankings of candidates or parties.
This approach is based on the theory of reflexive control
developed by Vladimir Lefebvre \citep{lef1,lef2}. This approach
yields a basis for typifying election campaigns. Basically, these
campaigns could be separated into three classes: the first class
includes the campaigns with high mobilization of voters, the
second class includes the campaigns with high electoral tension,
and the last one includes the campaigns of indifferent (or
proximal to that latter) choice of voters.

\section{Elections under indifferent choice of voters}\label{2}
Further, we shall consider the electoral processes (and its
peculiarities, in turn) of the Russian Federation. Russia has
bicameral parliament, with State Duma as the lower house, and the
Council of Federation as the upper chamber with the lower house
members elected through the mixed electoral system. That latter
implies that a half of the seats at the State Duma (225 seats of
total 450) are to be filled through the party list system, while
the other half is filled through the elections at the 225 single
constituencies. Here we shall consider both options.

Let's start from the simplest hypothetical situation. Suppose two
candidates race in a single constituency, say, to occupy a
position, which must be occupied through the elections process
(like the President of the Russian Federation, or a governor).
Suppose, then, that these two candidates are indistinguishable
persons from the voters point of view. Of course, such
hypothetical situation never occurs in real political campaign:
every candidate always tends to figure out his/her differences
from the rival candidates. Anyway, let us suggest at the first
step that two candidates do not differ, in the minds of voters,
and all the revisions and sharpening of our theory would be done
later, starting from that original point.

The essential question is what pattern of votes distribution one
should expect in such situation of a shortage of definiteness?
Probability theory makes one expect 50~$\!$\% to 50~$\!$\%
breakdown of votes. Human nature, nevertheless, fails to simulate
really random behaviour; this failure yields in a distortion of
the ratio of votes, according to the \textbf{golden section} law:
\begin{equation}\label{glsc}
0.618034\ldots \; \div \;\: 0.381966\ldots \;.
\end{equation}
Probably this fact was originally found by Vladimir Lefebre (see
\citep{lef2, lef3, lef1, G4, G5, AW}). Here one may expect to meet
the same behavioral pattern, when any differences between two
candidates seem to be negligible\footnote{We do not discuss the
way how one may head in such competition.}.

Actually, three candidates take part in a racing, not two. The
third candidate is not a physical person, rather it is an {\sl
``Against all candidates"} option in a ballot. A
relation~(\ref{glsc}) must be replaced, then, with
\begin{equation}\label{gl3}
0.50000\ldots \; \div \;\: 0.30902\ldots \; \div \;\:
0.19098\ldots \;.
\end{equation}
Here one might assume that the least value of (\ref{gl3})
corresponds to the option {\sl ``Against all candidates"}. This
assumption sounds natural, since a leading choice of that option
means a comparatively greater definiteness in an electoral
preference of voters, that it is supposed by the original
hypothesis. And, more generally, if $N$ candidates (or parties)
take part in the race, then the portions of ballots gathered by
the racing participants should be ranked proportionally to
Fibonacci numbers \citep{fib}. These are defined with the
recursion relation
\begin{equation}\label{chfib}
\phi_{i+2} = \phi_{i+1} + \phi_i \;{,} \
\end{equation}
with $\phi_1 = \phi_2 = 1$. A motivation behind this approach is a
premise that a person makes an indifferent choice in any
neighboring couple of candidates. A background theory for the
distorted choice in a case of two options is described in detail
by V.Lefebvre.

\section{Some cases of elections at Russia}\label{res}
Let now consider some specific cases of the elections run in
Russia. Both personal and party list elections would be
considered.
\subsection{Elections of the governor of Krasnoyarsk Krai, September 2002}\label{hlo}
The election of governor of Krasnoyarsk Krai took place in
September 2002. Two rounds have been carried out; Alexander
Khloponin (future winner) and Alexander Uhss advanced to the
second round. $46{,}82\!$~\% of all registered voters had taken
part in the elections. A winner (Alexander Khloponin) gathered
$48{,}07\!$~\% of votes, and Alexander Uhss gathered
$41{,}83\!$~\% of all votes; $9{,}15\!$~\% of all voters voted
against both candidates.

There was a high level of electoral (and political) tension during
these elections. It was evident from outer features, such as an
abundant drive, both outdoor, and via electronic mass-media one,
and from numerous scandals; the hottest one was with the approval
of the election results by the regional Election Commission. Also,
a number of staff involved in the campaign and various events was
very high, from both sides, for such type of racing.

\subsection{Election of Mayor of the city of Krasnoyarsk}\label{pim}
The Mayor elections took place in December of 1999. Three
candidates had taken part at these elections, including the
empowered major. There were one-round elections; i.e. the winner
(same person had been elected for the major position) gathered
$87{,}91\!$~\% of the votes, the option {\sl ``Against all
candidates"} had been chosen by $7{,}85\!$~\% of the voters who
entered the polling stations. Candidate came in the second
position gathered $3{,}42\!$~\% of the ballots. It should be
stressed, that the option {\sl ``Against all candidates"}, not the
real candidate, captured second rank.

These elections were characterized by the lowest possible
electoral tension. On the contrary, the elections to the city
Council exhibited a high level of the electoral tension
(simultaneously, the members of the city Council had been
elected). The major's block (that was an entity to be elected via
party list system) had gathered less than $25\!$~\%. Three issues
played the key role in the victory of Mr~Pimashkov (who won the
major position): first, he was the major already in the office
before the racing; second, he provided oppositon to governor Lebed
(who met with the minimal support of the residents of the city of
Krasnoyarsk); and, third, all other contenders were beyond any
criticism, frankly speaking.

\subsection{Election of Mayor of the city of Norilsk}\label{nor}
The elections of the major of the city of Norilsk took place in
October 2003. Several candidates have taken part in the elections,
which were carried out in a single round, with $34{,}2\!$~\% of
all registered voters coming to the polls. The winner
(Mr~Melnikov) gathered $51{,}34\!$~\% of the ballots, and the
secondly ranked contender gathered $33{,}21\!$~\% of all the
ballots; $15{,}45\!$~\% of all voters had chosen {\sl ``Against
all candidates"} option.

These results seem to be the closest to the theoretical situation
of a low definiteness of voters. Indeed, a high political tension
did not transform into an electoral one. The point is that the
result of the elections was the matter of interest of two powerful
financial groups, while the interests of voters felt off a
discussion of any political forces taking part in the elections.
Such situation appears to be rather close to an indifferent
choice, while the preference of voters was the matter of habit.
The winner connected well with his constituents.

\subsection{Preliminary survey of the cases}
It is evident, that the ranking observed in the races is far from
the theoretically expected. This fact is obvious: any real
campaign \textbf{must} differ from a situation of an indifferent
choice. Probably, the situation of the election of the major of
the city of Norilsk (see~\ref{nor}) looks the closest one to a
``reference" distribution. The data show that the preferences of
voters were rather clear and founded. It is a well known fact,
that a significant part of voters taking part in elections makes
their decision upon entering a polling station. It does not mean
that they choose indifferently; however, the key difference in
Western and up-to-date Russian electoral experience is the
difference of the background for a choice development. While
western voters tend to base their choice on a history of voting,
the Russian voters pay very much attention to attitude, mood and
feelings.

On the contrary, the case of election of the major of the city of
Norilsk shows that the situation of a low definiteness (or rather
proximate to that one) may occur, when public leaders or
politicians do not express their programme clearly. Such
conspiracy fails to present the public interests in the programs
of the candidates (see Section~\ref{nor}). The main political
players prominent in Norilsk, did not express their position
towards the presumptive victory of Mr~Melnikov (currently, the
mayor of the city). In particular, top managers of ``Norilsk
Nickel'' failed to say explicitly that they would not cooperate
with Mr~Melnikov, in case of his election. This conspiracy yielded
a wide spread confidence of the voters that the policy of the
corporation (and their well-being, which totally depends of the
corporation policy) would remain the same. In consequence, the
choice was irrational and was determined by emotional and
psychological reasons. Mr~Melnikov was emotionally closer to the
residents of the city.

\section{Party list system elections}\label{mnog}
Practically, one can never meet the ideal situation of indifferent
choice. Even experimental conditions could hardly provide such
situation. Following factors severely damage the original
assumption about the complete ignorance of voters and
indifferentness in their choice. A specificity of human mind
procedures of multivariant classification is the first factor.
This specificity may manifest in the occurrence of stable and
long-term preferences towards the entities within the list, but
not for any couple of them. Say, a voter always prefers ``Union of
Right Forces" for ``Yabloko", but fails to place this (strictly
ordered) couple within a line of other parties and entities. An
impact of the external conditions (advertising, political
sketches, etc.) is the second factor. This factor may manifest
itself through the specific electoral technologies, ranging from
activity of an entity's double (or a person's double) to the
dissemination by a candidate of the negative information
concerning himself, or herself. Such information may be not
appreciated by the voters thus yielding a growth in the
candidate's rating.

Let's now consider some results of the elections carried out on a
party list system. The results are shown in charts, where the left
axis represent a real rating of an entity, and the right one shows
the \textbf{deviation index} between real and theoretically
expected (\ref{chfib}) rating observed for a completely
indifferent choice. The excess of the index over $1$ means that
the entity gathered more ballots, than may be expected for a
completely indifferent choice\footnote{It should be stressed, we
do not discuss the issue of a leadership in such line of
entities.}.

\subsection{Elections of Legislative Assembly of Krasnoyarsk Krai, 2001}\label{krai}
\begin{figure}[ht]
\begin{center}
\includegraphics{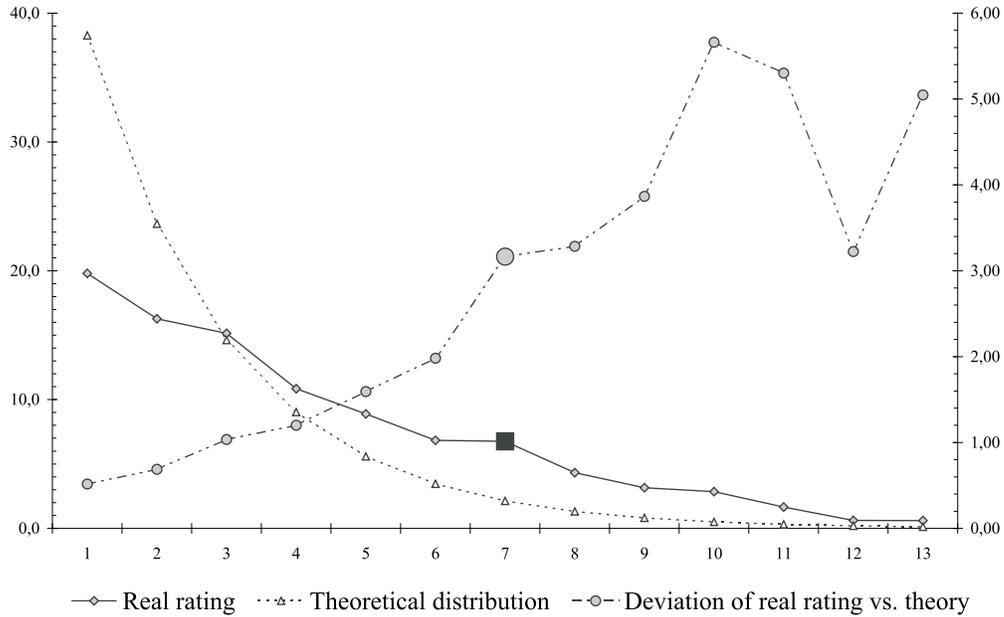}
\end{center} \caption{Legislative Assembly of Krasnoyarsk krai elections in  2001.
Left axis represents the rating of political entities, the right
one shows the deviation between real rating and theoretical
distribution. Greater label shows {\sl ``Against all entities"}
option.}\label{f2}
\end{figure}

Figure~\ref{f2} shows the results of elections of the Legislative
Assembly of Krasnoyarsk Krai, which took place in 2001. The
pattern in this figure is rather far from the one expected for the
situation of completely indifferent choice.
It looks very natural, since the elections were running in the
highly polarized political situation, where two coalitions
competed relentlessly for the seats in the Assembly. The extended
markers point out the portion of {\sl ``Against all entities"}
option. Following are the entities, as they are enumerated in the
figure: 1 stands of ``Ours!'' coalition; 2 stands for coalition
for Anatoly Bykov; 3 stands for ``For Lebed!'' coalition; 4 stands
for the Communist party of the Russian Federation; 5 stands for
``Medved" coalition; 6 stands for the Northern party; 7 stands for
{\sl \textbf{``Against all entities"}}; 8 stands for ``Hope and
buttress'' coalition; 9 stands for Zubov's coalition; 10 stands
for ``Communists and patriots of Krasnoyarsk motherland for the
soviet power'' coalition; 11 stands for Union of Right Forces; 12
stands for Liberal Democratic Party; and 13 stands for ``Yabloko".

Theoretically, if one cumulates the ballots gathered by two
opposing coalitions (that are ``Ours!'' coalition Bykov's
coalition) into a single pool, then the deviation index ($0,94$)
becomes very close to $1$. This fact shows that any splitting of
political players under the strong electoral tension would yield a
decrease of the ballot portion which might be gathered by a leader
under the indifferent choice. Surprisingly, these elections show a
non-monotonous behaviour of the deviation index.

\subsection{State Duma elections}\label{duma}
There were three elections to the State Duma since 1993. State
Duma has been reinstated in 1993, just after the dismissal of the
Supreme Council of the Russian Federation. Here we analyze the
results of the elections of 1999 and 2003. Below are two lists of
parties and associations, which had taken part at these elections;
the entities are enumerated according to the ballot-paper.

There were 26 entities taking part in the elections of 1999:
Conservative movement of Russia is number 1; Russian nationwide
union is number 2; ``Women of Russia" movement is number 3; Stalin
block for USSR is number 4; ``Yabloko" party is number 5;
``Communists and work people for the Soviet Union" movement is
number 6; ``Peace. Labor. May." movement is number 7; Block of
Nikolaev and Fyodorov is number 8; ``Intellectual heritage"
national movement is number 9; Congress of Russian communes is
number 10; Peace and Unity party is number 11; Russian party of
women defense is number 12; Medved (later ``United Russia" party)
is number 13; ``Social democrats" movement is number 14;
Nationwide movement for the army support is number 15; Zhirinovsky
block (later Liberal Democrats party) is number 16; ``For civic
dignity" movement is number 17; ``Motherland~--~All Russia" is
number 18; Communist party of Russian Federation is number 19;
``Russian line" movement is number 20; All-Russian political party
of peoples is number 21; Union of Right Forces is number 22;
``Russia -- our Motherland" is number 23; Socialistic party of
Russia is number 24; Party of pensioners is number 25; Russian
socialistic party is number 26; and {\sl Against all entities}
option is number 27.

Twenty three entities took part in the elections of 2003: ``Unity"
movement is number 1; Union of Right Forces is number 2; Party of
pensioners is number 3; ``Yabloko" party is number 4; ``For saint
Russia" is number 5; Joint party ``Russia" is number 6; ``New line
--- Russia on Car" is number 7;  Republican party is number 8;
``Greenpeace" is number 9; Party of farmers is number 10; Real
patriots of Russia is number 11; All-Russian political party of
peoples is number 12; Party of constitutional democrats is number
13; ``Great Russia -- Euroasian Union" block is number 14; ``SLON"
party is number 15; ``Fartherland" block is number 16;  Peace and
Unity party is number 17; Liberal Democrats party (former
Zhirinovsky block) is number 18: Russian party of life is number
19; ``United Russia" party (former Medved) is number 20;
Democratic party of Russia is number 21; Party of business
development is number 22; Communist party of Russian Federation is
number 23, and finally, number 24 stands for {\sl Against all
entities"} option.

Both lists of parties look rather long and very pretentious.
Twenty six and twenty three entities took part in the elections of
1999 and 2003, respectively. Only nine parties and associations
had taken part in both elections, with respect to the
transformations and integrations of smaller entities. This patters
is an evidence of high instability of political entities in
Russia. Probably two or three more entities from the list of
the elections of 2003 could be derived from some entities from the
list of the elections of 1999, while formally they seem to be
independent.

\subsubsection{Elections of 1999}\label{d99}
\begin{figure}[ht]
\includegraphics{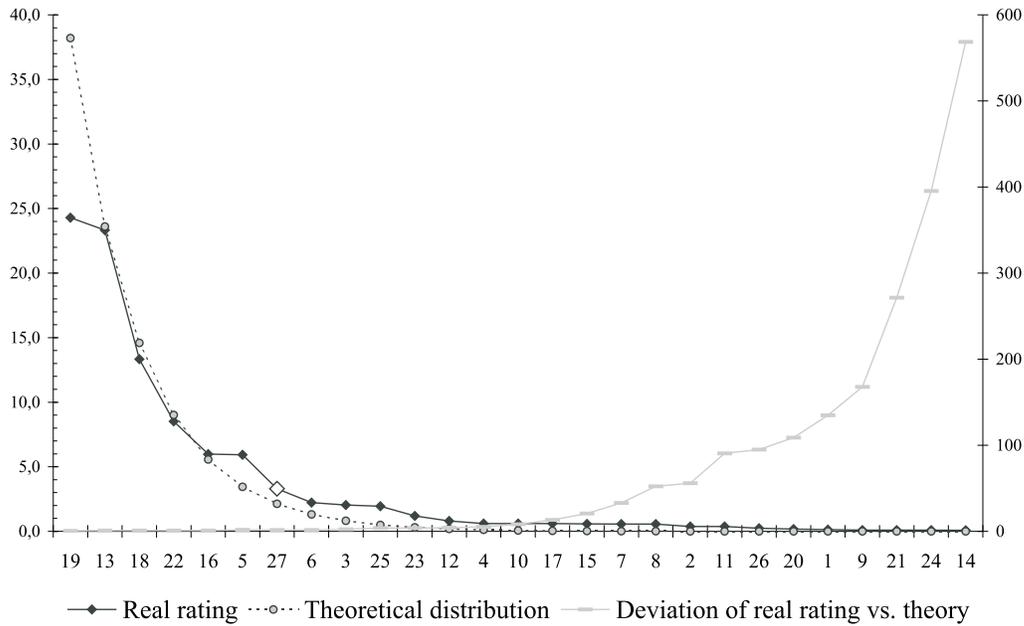}
\caption{State Duma elections of  1999. Left axis shows the
party's rating, but the right one represents the \textbf{deviation
index} between real and theoretical rating. The parties and
entities are ranked according to the ballot; see details in the
text.}\label{f3}
\end{figure}

Six parties had overcome the legislatively established threshold
of $5\!$~\% ballots, to be represented in the State Duma. The
pattern of ballot distribution differs significantly from the one
observed for the Legislative Assembly of Krasnoyarsk Krai
elections (see above section~\ref{krai}). Figure~\ref{f3} shows
the results of the State Duma elections in 1999.
A number in the horizontal axis in Figure~\ref{f3} corresponds to
the entities as they listed in a ballot and ranked in descending
order with respect to the votes gathered by a party.

\subsubsection{Elections of 2003}\label{d3}
\begin{figure}[ht]
\includegraphics{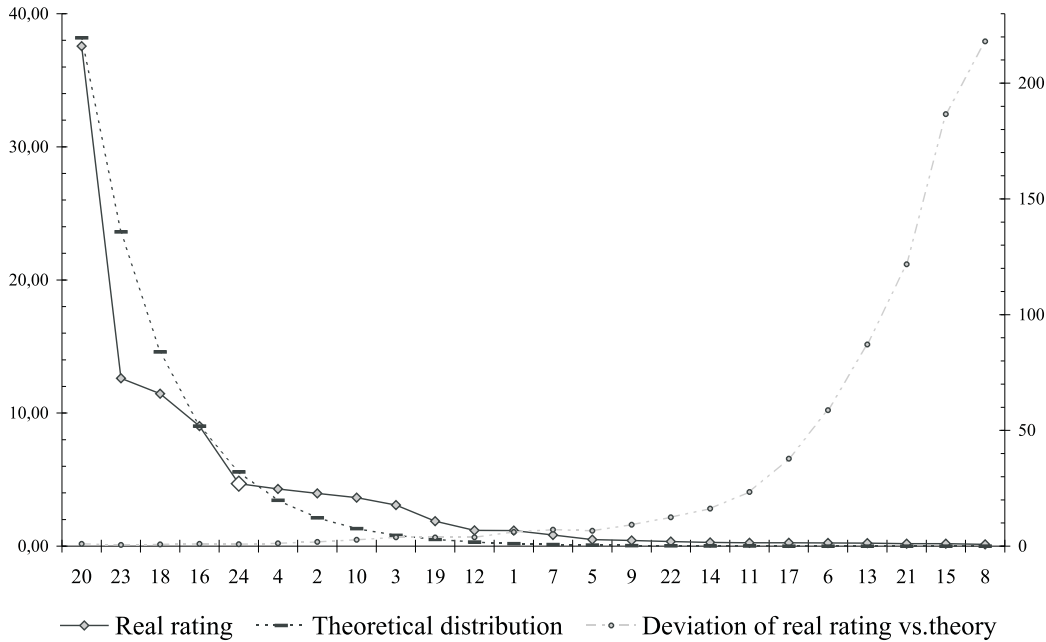}
\caption{State Duma elections of 2003. Left axis shows the party's
rating, but the right one represents the \textbf{deviation index}
between real and theoretical rating. The parties and entities are
ranked according to the ballot; see details in the
text.}\label{ris3}
\end{figure}

Figure~\ref{ris3} shows the results of the elections of 2003. The
parties (or associations) in this figure are ranked in descending
order according to their rating. Again, the enlarged label on the
chart indicates the {\sl ``Against all entities"} option.
This picture reveals a significant difference of the elections
from those of 1999. Firstly, those former show much more
concordant pattern of the observed ballot distribution, than the
theoretical one (see (\ref{chfib})). It means that these elections
were running in a situation of significantly less definiteness, in
comparison to the previous ones. Secondly, the deviation index
exhibits a three-fold decay. Thirdly, a two-fold growth of the
ballots given {\sl Against all entities} is manifested. We believe
that this manifestation corroborates the increased trend to vote
``by guess-work". These data show that all the minor entities that
fail to pass $5\!$~\% threshold nevertheless managed to pick up
much greater number of ballots, in comparison to the situation of
completely indifferent choice.

\section{Discussion and Conclusion}\label{disc}
Basic purpose of this paper is to develop a tool to classify
electoral campaigns from the point of view of the rationality vs.
irrationality of the voter's choice. The key issue is the question
of a definition of ``equilibrium" or ``indifferent" voting. That
latter is the voting where a voter fails to figure out his (or
her) rational arguments for choosing a specific candidate of a
specific party from the list. We believe that the outcome of the
elections under such circumstances should follow the proportion
described by Fibonacci numbers \citep{fib}.

One should distinguish such \textbf{``by guess-work" voting} from
the really random one. That latter should be interpreted according
to the probabilities theory. A voter fails to avoid a reflexive
pattern in the behavior, so an indifferent choice would exhibit a
bias in ballot distribution, among two options. The bias is
defined by the golden section law. Such bias in preferences is
peculiar for a human being, and constitutes a core feature of
human psychology \citep{G4,G5,AW}.

Any real electoral campaign would obviously differ from the
indifferent choice situation. This fact is evident due to the
behaviour of the deviation index which compares real rating and
theoretically expected one (\ref{chfib}). Roughly, one can compare
any electoral campaign to the reference situation of the
indifferent choice. There are two poles around this ``zero point"
pattern. The former is the situation of high mobilization of
voters (voters become the supporters), and the latter is the
situation of increased political (or electoral) tension. The
mobilization is manifested through the strong leadership of a
candidate, in comparison to the leadership with indifferent choice
of voters. High electoral tension manifests through the closer
ballots gathered by two competitors. Such proximity of ballots
means an occurrence of practically equal pools of supporters of
each contestant. Another reason of a discrepancy between golden
section theory and real vote distribution is a diversity of
people. We believe, the people differ among themselves, from the
point of view of their attitude to political issues; it may result
in permanent occurrence of a ``tail" in the deviation index.

No one real electoral campaign could be considered as the
indifferent choice situation. This latter is just a theoretical
issue which introduces a reference point for the analysis of
efficiency and public processes underlying the campaign.
Evidently, the societies with developed democracies never meet the
situation close to the indifferent choice. Probably, such pattern
of public attitude to politicians and parties could be observed in
transitional societies with lack of experience in democratic
institutions. Nevertheless, both practitioners and researchers of
that type of societies may capitalize on the campaign typifying
developed above.

\end{document}